\documentclass[pra,preprint,superscriptaddress,amsmath,amssymb,showpacs]{revtex4}
\usepackage{bm}
\usepackage{amsmath}
\usepackage[dvips]{graphicx}
\begin{document}

\title{Delbr\"uck scattering in combined Coulomb and laser fields}

\author{A. \surname{Di Piazza}}
\email{dipiazza@mpi-hd.mpg.de}
\affiliation{Max-Planck-Institut f\"ur Kernphysik, 69117 Heidelberg, Germany}
\author{A. I. \surname{Milstein}}
\email{milstein@inp.nsk.su}
\affiliation{Max-Planck-Institut f\"ur Kernphysik, 69117 Heidelberg, Germany}
\affiliation{Budker Institute of Nuclear Physics, 630090 Novosibirsk, Russia}

\date{\today}

\begin{abstract}
We study Delbr\"uck scattering in a Coulomb field in the presence of a laser field. The amplitudes are calculated 
in the Born approximation with respect to the Coulomb field and exactly in the parameters of the laser field
having arbitrary strength, spectral content and polarization.
The  case of high energy initial photon energy is investigated in detail for a monochromatic circularly polarized laser field. It is shown that the angular distribution of the process substantially differs from that for Delbr\"uck scattering in a pure Coulomb field. The value of the cross section under discussion may exceed the latter at realistic laser parameters that essentially simplify the possibility of the experimental observation of the phenomenon. The effect of high order terms in the quantum intensity parameter $\chi$ of the laser field is found to be very important already at relatively small $\chi$.
\end{abstract}
 
\maketitle
\section{Introduction}
The theoretical and experimental investigation of nonlinear Quantum Electrodynamics (QED) effects arising due to vacuum polarization in the presence of a strong external field renders possible testing QED when the nonperturbative effects in the field strength (contributions of higher-order terms) are very significant. One of these effects, the scattering of a photon in the electric field of atoms (Delbr\"uck scattering), has been investigated in detail both theoretically and experimentally \cite{PM75,MS94,GS95,Delbexp}. Also, a decisive progress has been obtained in the understanding of the process of photon splitting in an atomic electric field  \cite{LMS,LMMST03}. In \cite{splitexp} the first successful experimental observation of photon splitting in an atomic field is reported. As a result, it has been observed that higher orders of the perturbation theory with respect to the parameter $Z\alpha$ play an important role and essentially modify the cross sections of Delbr\"uck scattering and photon splitting in an atomic field at high photon energy (here, $Z$ is the nuclear charge number, $\alpha=e^2=1/137$ is the fine-structure constant, with $e$ being the electron charge and the system of units $\hbar=c=1$ is used). 

Besides the atomic field, photon splitting has also been investigated in other configurations of the external field, namely in constant electromagnetic fields \cite{Adler70,BB70,Adler71,Papanyan74,BMS869697} and in a laser field \cite{DMK2007}. In \cite{DMK2007} photon splitting in a laser field with arbitrary strength and frequency content has been investigated for an arbitrary energy of the incoming photon.  The photon splitting amplitude in a laser field depends on two parameters that, in the frame where the laser wave propagates in the direction anti-parallel to the momentum $\bm k_1$
of the initial photon, are given by
\begin{equation}\label{chieta}
\eta=\frac{\omega_1\omega_0}{m^2}\, ,\quad \chi=\frac{\omega_1}{m}\frac{E}{E_c}\,.
\end{equation}
Here, $\omega_1$ is the energy of the incoming photon, $\omega_0$ and $E$ are the characteristic frequency and electric field strength of the laser wave, $E_c=m^2/|e|=1.3\times 10^{16}\,$V/cm is the so-called critical electric field and $m$ is the electron mass. At present, the parameters of optical laser fields as well as of photon sources permit achieving nowadays values of $\chi\sim 1$ where non-perturbative effects in the laser field become important \cite{Emax}. We emphasize, however, that the structure of the photon splitting cross section in the case of both constant and laser fields does not allow one measuring this effect without considerable effort.

In the present paper, we consider a process which is closely related to photon splitting in a laser field. Namely, we study photon scattering (Delbr\"uck scattering) in combined Coulomb and laser fields. We perform our calculation in the Born approximation with respect to the Coulomb field and exactly in the parameters of the laser field. Thus, the amplitude obtained is linear in $Z\alpha$ and corresponds to photon splitting in a laser field when an initial real photon with the momentum $k^{\mu}_1=(\omega_1,\,\bm k_1)$, $k_1^2=0$, splits into a real photon with the momentum $k^{\mu}_2=(\omega_2,\,\bm k_2)$,  $k_2^2=0$, and a virtual photon (Coulomb quantum) with the momentum $k^{\mu}_3=(0,\,\bm q)$, $k_3^2=-{\bm q}^2$. We consider the most interesting case where a nucleus (source of the Coulomb field) is at rest and a plane wave, corresponding to the laser field, propagates in the direction $\bm\varkappa$ anti-parallel to the momentum $\bm k_1$. The plane wave is described by an arbitrary four-vector potential $A^{\mu}(\phi)=(0,\bm A(\phi))$, where $\phi=\varkappa x$, with $\varkappa^{\mu}=(1,\bm\varkappa)$, $\varkappa^2=0$, and $\bm\varkappa\cdot\bm A(\phi)=0$. The amplitudes are derived following the method described in detail in our paper \cite{DMK2007} and based on the operator technique developed in \cite{BMS75, BKMS75} for the calculation of the polarization and mass operators in a laser field. The advantage of the operator technique is that it enables one to calculate the amplitudes of the processes without employing the explicit form of the electron propagator in the external field. We first derive the explicit expressions of the  amplitudes for a general vector potential of the form $\bm A(\phi)$ (Section II). Then, we consider the case of a monochromatic laser field with circular polarization (Section III) and analyze in detail the properties of the amplitudes for the most interesting parameter region: $\omega_0\ll m$ and $\omega_1\gg m$ (Section IV). Finally, in Section V, the results are discussed and the main conclusions of the paper are presented.

\section{General form of the amplitudes}

\begin{figure}[ht]
\begin{center}
\includegraphics[width=10cm]{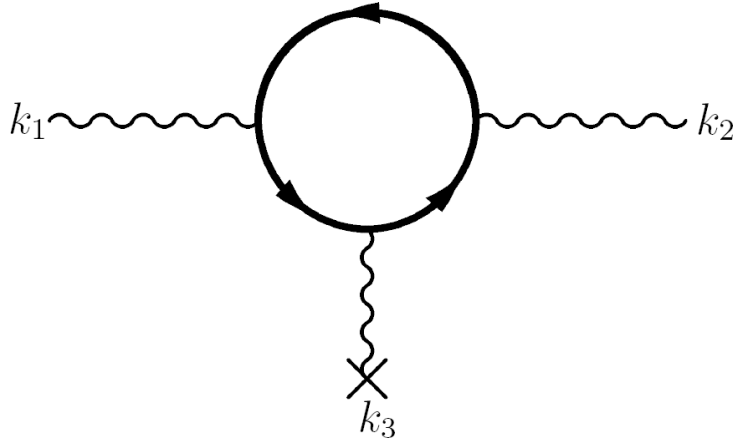}
\caption{\label{PSD} Feynman diagram for the Delbr\"uck scattering amplitude in combined Coulomb and laser fields
calculated in the Born approximation with respect to the Coulomb field. This diagram corresponds to the Furry representation and the thick line denotes the electron propagator (Green's function) in the laser field. The wavy lines symbolize the external photons, $k^{\mu}_1$ and $k^{\mu}_2$ are the momenta of the initial and final photons, respectively, $k^{\mu}_3$ is the momentum of the Coulomb quantum. The diagram with the permutation $k^{\mu}_2\leftrightarrow k^{\mu}_3 $ has to be added.} 
\end{center} 
\end{figure}

The amplitude $M$ of  Delbr\"uck scattering in combined Coulomb and laser fields calculated in the Born approximation with respect to the Coulomb field  is represented by the Feynman diagram shown in Fig. \ref{PSD}. It has the form
\begin{equation}\label{tensorM}
M=-\frac{(4\pi)^2(Ze)}{{\bm q}^2}{\cal M}^{\mu\nu\rho}e_{1\mu}e_{2\nu}^*e_{3\rho}^*\,,
\end{equation}
where $e_{1\mu}$ and $e_{2\mu}$ are  the initial and final photon polarization four-vectors, respectively, 
 $e_{3\mu}=g_{\mu 0}$ is  the polarization four-vector of the Coulomb quantum, and the formal expression for 
the tensor ${\cal M}^{\mu\nu\rho}$ in the Furry representation reads
\begin{eqnarray}\label{MG}
{\cal M}^{\mu\nu\rho}&=&ie^3\int\,d^4 x\,\mbox{Tr}\langle x|
\mbox{e}^{-ik_1x}\,\gamma^\mu\frac{1}{\hat{P}-m+i0}\mbox{e}^{ik_2x}\,\gamma^\nu\frac{1}{\hat{P}-m+i0}
\mbox{e}^{ik_3x}\,\gamma^\rho\frac{1}{\hat{P}-m+i0}|x\rangle\,\nonumber\\
&&+(k^{\mu}_2\leftrightarrow k^{\mu}_3\, , \,\nu\leftrightarrow \rho)\, ,
\end{eqnarray}
Here $\hat { P}={ P}_\mu\gamma^\mu$, with $P_\mu=i\partial_\mu-eA_\mu(\phi)$ and $\gamma^\mu$ being the Dirac matrices.

By using gauge invariance it is possible to simplify the calculation of the amplitude $M$. To that purpose we introduce for each photon with the four-momentum  $k^\mu_i$ ($i=1,2,3$) two four-vectors $\varepsilon_{i,\lambda_i}^\mu$, ($\lambda_i=1,2$)
\begin{eqnarray}\label{varepsilon}
&&\varepsilon_{i,\lambda_i}^\mu=a_{\lambda_i}^\mu-\frac{k_ia_{\lambda_i}}{\varkappa k_i}
\varkappa^\mu\, ,\quad  \varepsilon_{i,\lambda_i}\varepsilon_{j,\lambda_j}= -\delta_{\lambda_i,\lambda_j}\,\nonumber\\
&&a_\lambda^0=0\,,\quad ({\bm a}_\lambda)^2=1\, ,\quad {\bm a}_1\cdot{\bm
a}_2=0\,,\quad {\bm a}_\lambda\cdot {\bm\varkappa}=0.
\end{eqnarray}
For the virtual photon (Coulomb quantum), we also introduce the third four-vector
\begin{eqnarray}\label{varepsilon3}
\varepsilon_{3,3}^\mu=\sqrt{\frac{k_3^2}{(\varkappa k_3)^2}}\,
\left(\varkappa^\mu-\frac{(\varkappa k_3)}{k_3^2}k_3^\mu\right)\, ,\quad \varepsilon_{3,3}\varepsilon_{3,3}=-1
\end{eqnarray}
which is perpendicular to $k_3^{\mu}$, $\varepsilon^{\mu}_{3,1}$, and $\varepsilon^{\mu}_{3,2}$ (see Eq. (\ref{varepsilon})). Then, gauge invariance allows us to write the tensor ${\cal M}^{\mu\nu\rho}$ introduced in Eq. (\ref{tensorM}) in the form:
\begin{equation}\label{expandM}
{\cal M}^{\mu\nu\rho}=-\sum_{\substack{\lambda_1,\lambda_2=1,2\\ \lambda_3=1,2,3}}
R_{\lambda_1\lambda_2\lambda_3}\varepsilon_{1,\lambda_1}^\mu
\varepsilon_{2,\lambda_2}^\nu\varepsilon_{3,\lambda_3}^\rho \, .
\end{equation}
From this equation we have
\begin{equation}\label{RVM}
R_{\lambda_1\lambda_2\lambda_3}={\cal M}_{\mu\nu\rho}
\varepsilon_{1,\lambda_1}^\mu
\varepsilon_{2,\lambda_2}^\nu\varepsilon_{3,\lambda_3}^\rho \, .
\end{equation}
By employing the coefficients $R_{\lambda_1\lambda_2\lambda_3}$, we write the amplitude $M$ as
\begin{equation}\label{MviaS}
M=\frac{(4\pi)^2(Ze)}{{\bm q}^2}\sum_{\substack{\lambda_1,\lambda_2=1,2\\ \lambda_3=1,2,3}}
R_{\lambda_1\lambda_2\lambda_3}(\varepsilon_{1,\lambda_1}e_1)
(\varepsilon_{2,\lambda_2}e_2^*)(\varepsilon_{3,\lambda_3}e_3^*) \, .
\end{equation}
In this equation it is not necessary to perform any subtractions since it automatically fulfills gauge invariance.

It is convenient to perform the calculations in terms of the helicity amplitudes $M_{\sigma_1\sigma_2}$, with $\sigma_j=\pm 1$. In this case, the polarization vector $\bm e_{\sigma}$ of each external real photon with the wave vector ${\bm k}$ and the energy $\omega$ satisfies the relations $\bm e_\sigma\cdot\bm k=0$ and $\bm e_{\sigma}\times\bm k=i\sigma\omega \bm e_\sigma$. Due to momentum conservation in the laser field, the final transverse momenta $\bm k_{2\perp}$ and $\bm k_{3\perp}$ (with respect to the laser propagation direction) are equal in modulus and opposite in direction:
$\bm k_{2\perp} = -\bm k_{3\perp}$ and $\bm{k}_{2\perp}\cdot\bm\varkappa=\bm{k}_{2\perp}\cdot\bm k_1=0$. We direct the vector $\bm a_1$ along the vector $\bm{k}_{2\perp}$ and set $\bm a_2$ to be $\bm a_2=\bm\varkappa\times\bm a_1$, then
\begin{eqnarray}
\bm e_{1\,,\sigma_1}&=&-\frac{1}{\sqrt{2}}(\bm a_2+i\sigma_1\bm a_1)\, ,\nonumber\\
\bm e_{2\,,\sigma_2}&=&\frac{1}{\sqrt{2}}(\bm a_2-i\sigma_2 \bm a_2\times\bm k_2/\omega_2)\, .
\end{eqnarray}
The  polarization four-vectors $e^{\mu}_{i,\sigma_i}=(0,\bm e_{i\,,\sigma_i})$ and $e^{\mu}_3$ have the following products with the four-vectors $\varepsilon^{\mu}_{i,\lambda_i}$:
\begin{eqnarray}
&&\varepsilon_{1,\lambda_1}e_{1,\sigma_1}=\frac{1}{\sqrt{2}}(i\sigma_1\delta_{\lambda_1,1}+\delta_{\lambda_1,2})\, ,\nonumber\\
&&\varepsilon_{2,\lambda_2}e_{2,\sigma_2}^*=\frac{1}{\sqrt{2}}(i\sigma_2\delta_{\lambda_2,1}-\delta_{\lambda_2,2})\, ,\nonumber\\
&&\varepsilon_{3,\lambda_3}e_{3}^*=\varepsilon_{3,\lambda_3}^0=
-\frac{|\bm{k}_{2\perp}|}{\varkappa k_3}\,\delta_{\lambda_3,1}+
  \sqrt{\frac{k_3^2}{(\varkappa k_3)^2}}\,\delta_{\lambda_3,3} \, .
\end{eqnarray}

Using these relations we obtain from Eq. (\ref{MviaS})
\begin{equation}\label{MviaS1}
M_{\sigma_1\sigma_2}=\frac{(4\pi)^2(Ze)}{2{\bm q}^2}
\left(\sigma_1\sigma_2{\cal R}_{11}+{\cal R}_{22}+i\sigma_1{\cal R}_{12}-i\sigma_2{\cal R}_{21}\right)\, ,
\end{equation}
where
\begin{equation}\label{MviaS2}
{\cal R}_{\lambda_1\lambda_2}=\frac{|\bm{k}_{2\perp}|}{\varkappa k_3}
\left(R_{\lambda_1\lambda_2\,1}^{(2)}+R_{\lambda_1\lambda_2\,1}^{(3)}\right)-
\sqrt{\frac{k_3^2}{(\varkappa k_3)^2}}R_{\lambda_1\lambda_2\, 3}^{(3)}\, .
\end{equation}
The calculation of both terms, $R_{\lambda_1\lambda_2\lambda_3}^{(2)}$ and $R_{\lambda_1\lambda_2\lambda_3}^{(3)}$,
is quite similar to the calculation performed in \cite{DMK2007} and we present here only the final results.  For the term  $R_{\lambda_1\lambda_2\lambda_3}^{(2)}$ with $\lambda_3=1,2$, we obtain 
\begin{eqnarray}\label{S2final}
&&R_{\lambda_1\lambda_2\lambda_3}^{(2)}=
-\frac{me^3}{(2\pi)^2}\sum_{j=1,2,3}\delta_{(j)}\iint\limits_0^{\quad\infty}\frac{ds_1\,ds_2}{(s_1+s_2)^2}
\exp\left[-i(s_1+s_2)-i\frac{s_1s_2}{s_1+s_2}Q^2\delta_{j,3}\right]\nonumber\\
&&\times\int d\phi\exp\left\{i\Omega\phi
+i\frac{4\omega_1\Omega}{m^2}\varphi_j-i(s_1+s_2)\left[\int_0^1dy{\bm\Delta}^2(y\varphi_j)-
\left(\int_0^1dy{\bm\Delta}(y\varphi_j)\right)^2\right]\right\}\nonumber\\
&&\times\left[-2{\bm a}_{\lambda_j}\cdot\int_0^1dy
\bm{\Delta}(y\varphi_j)+\frac{(s_1+s_2)^2}{2s_1s_2}{\bm
a}_{\lambda_j}\cdot\bm{\Delta}(\varphi_j)\right]\, .
\end{eqnarray}
In this expression we introduced the quantities
\begin{eqnarray}\label{S2notation}
&&\delta_{(1)}=\delta_{\lambda_2,\lambda_3}\, ,\quad \delta_{(2)}=\delta_{\lambda_1,\lambda_3}\, ,\quad
\delta_{(3)}=\delta_{\lambda_1,\lambda_2}\, ,\quad \nu_1=-1\, ,\quad  \nu_{2,3}=\frac{\varkappa k_{2,3}}{\varkappa k_1}\,, \quad Q^2=\frac{\bm{q}^2}{m^2}\,,
\nonumber\\
&&\varphi_j=-\frac{s_1s_2}{s_1+s_2}\nu_j\,,
\quad {\bm\Delta}(u)=\frac{e}{m}[{\bm A}(\phi+4u\omega_1/m^2 )- {\bm A}(\phi)]\,, \quad \Omega=\omega_2- \omega_1\, .
\end{eqnarray}
Note that $R_{\lambda_1\lambda_2\lambda_3}^{(2)}$ contains only odd powers of the external field.

We pass now to the term $R_{\lambda_1\lambda_2\lambda_3}^{(3)}$. In order to write it in a more compact form, we introduce the convenient notation
\begin{eqnarray}
&&\quad S=s_1+s_2+s_3\, ,\,\, \tau_1= \frac{s_2\nu_2-s_3\nu_3}{ S}\,,\,\,
\tau_2=\nu_2-\tau_1\,,\quad
\tau_3=\nu_3+\tau_1\,,\,
\nonumber\\
&& \bm{ D}_1(y)= {\bm \Delta}(s_2\tau_2y+s_3\tau_3(1-y))\,, \,
\bm{ D}_2(y)= {\bm \Delta}(s_2\tau_2y)\,, \,
\bm{ D}_3(y)= {\bm \Delta}(s_3\tau_3y)\,,\nonumber\\
&&{\bm F}=\frac{1}{ S}\int_0^1dy[s_1\bm{ D}_1(y)+s_2\bm{
D}_2(y)+s_3\bm{ D}_3(y)]\,
,\nonumber\\
&&F_0=  S{\bm F}^2-\int_0^1dy[s_1\bm{ D}_1^2(y)+s_2\bm{
D}_2^2(y)+s_3\bm{ D}_3^2(y)]\, ,\nonumber\\
&& \bm G=\frac{2s_1}{ S}\int_0^1dy\,[s_2(\bm{ D}_1(y)-\bm{
D}_2(y))+s_3(\bm{ D}_1(y)-\bm{ D}_3(y))]\, ,\nonumber\\
&&\bm \rho=\frac{\bm{k}_{2\perp}}{m}\, ,\quad {\bm N}_1={\bm F}-\frac{s_1\bm \rho}{ S}\,,\quad
{\bm N}_2=\bm F-\bm{
D}_2(1)+\frac{s_3\bm \rho}{ S \nu_2}\,,\nonumber\\
&&{\bm N}_3={\bm
F}-\bm{ D}_3(1)+\frac{s_2\bm \rho}{ S \nu_3}
\,,\quad \bm V=\frac{1}{2}(\bm d_1+\bm d_2+\bm d_3)\,,\nonumber\\
&&\bm d_1=\frac{\bm D_2(1)-\bm D_3(1)}{2\tau_1}\, ,\quad
\bm d_2=\frac{\bm D_2(1)}{2\tau_2}\, ,\quad
\bm d_3=\frac{\bm D_3(1)}{2\tau_3}\,.
\end{eqnarray}
We obtain for $\lambda_3=1,2$:
\begin{eqnarray}\label{S3final}
&&R_{\lambda_1\lambda_2\lambda_3}^{(3)}=\frac{ime^3}{2\pi^2}{\cal P}\int\negthickspace\negthickspace\int\limits_0^{\infty}
\negthickspace\negthickspace\int\frac{ds_1\,ds_2\,ds_3}{ S^2}\int d\phi\nonumber\\
&&\times\exp\left[i\left(\Omega\phi+\frac{4\omega_1\Omega s_2s_3}{m^2 S}-
\frac{s_1s_3}{S}Q^2+F_0- S +{\bm \rho}\cdot{\bm G}
\right)\right]\nonumber\\
&&\times\Bigg<-4({\bm a}_{\lambda_1}\cdot{\bm N}_1) ({\bm a}_{\lambda_2}\cdot{\bm N}_2)
({\bm a}_{\lambda_3}\cdot{\bm N}_3)+4({\bm a}_{\lambda_2}\cdot{\bm N}_2)
 ({\bm a}_{\lambda_3}\cdot{\bm N}_3)({\bm a}_{\lambda_1}\cdot{\bm V})
\nonumber\\
&&-4\nu_3({\bm a}_{\lambda_1}\cdot{\bm N}_1) ({\bm a}_{\lambda_2}\cdot{\bm N}_2)
({\bm a}_{\lambda_3}\cdot{\bm V})-4\nu_2({\bm a}_{\lambda_1}\cdot{\bm N}_1)
({\bm a}_{\lambda_3}\cdot{\bm N}_3)({\bm a}_{\lambda_2}\cdot{\bm V})
\nonumber\\
&&+({\bm a}_{\lambda_1}\cdot{\bm N}_1)\bigg\{
\frac{[\bm \rho\cdot\bm \rho]_{\lambda_2\lambda_3}}{2\nu_2\nu_3}+
\frac{\nu_2Q^2}{2\nu_3}\delta_{(1)}+
[{\bm \rho}\cdot
(\bm d_1-\bm d_2-\bm d_3 )]_{\lambda_2\lambda_3}-2\nu_2\nu_3[\bm d_1\cdot(\bm d_2+\bm d_3)]_{\lambda_2\lambda_3}\bigg\}
\nonumber\\
&&-({\bm a}_{\lambda_2}\cdot{\bm N}_2)\bigg\{
\frac{[\bm \rho\cdot\bm \rho]_{\lambda_1\lambda_3}}{2\nu_3}+
\frac{Q^2}{2\nu_3}\delta_{(2)}+[{\bm \rho}\cdot
(\bm d_1+\bm d_2-\bm d_3 )]_{\lambda_1\lambda_3}-2\nu_3[\bm d_3\cdot(\bm d_1+\bm d_2)]_{\lambda_1\lambda_3}\bigg\} \nonumber\\
&&-({\bm a}_{\lambda_3}\cdot{\bm N}_3)\bigg\{
\frac{[\bm \rho\cdot\bm \rho]_{\lambda_1\lambda_2}}{2\nu_2}+[{\bm \rho}\cdot
(\bm d_1-\bm d_2+\bm d_3) ]_{\lambda_1\lambda_2}-2\nu_2[\bm d_2\cdot(\bm d_1+\bm d_3)]_{\lambda_1\lambda_2}\bigg\}
\nonumber\\
&&+({\bm a}_{\lambda_1}\cdot{\bm \rho})[{\bm \rho}
\cdot\bm d_1]_{\lambda_2\lambda_3}-\frac{1}{\nu_2}({\bm a}_{\lambda_2}
\cdot{\bm \rho})[{\bm \rho}\cdot\bm d_3]_{\lambda_1\lambda_3}-
\frac{1}{\nu_3}({\bm a}_{\lambda_3}
\cdot{\bm \rho})[{\bm \rho}\cdot\bm d_2]_{\lambda_1\lambda_2}\nonumber\\
&&-\frac{\rho^2}{2}\bigg\{[{\bm a}_{\lambda_1}\cdot\bm d_1]_{\lambda_2\lambda_3}
-\frac{1}{\nu_2}[{\bm a}_{\lambda_2}\cdot\bm d_3]_{\lambda_1\lambda_3}
-\frac{1}{\nu_3}[{\bm a}_{\lambda_3}\cdot\bm d_2]_{\lambda_1\lambda_2}
\bigg\}-\frac{\nu_2Q^2}{2\nu_3}[{\bm a}_{\lambda_3}\cdot\bm d_2]_{\lambda_1\lambda_2}\nonumber\\
&&+\frac{1}{\nu_2\nu_3}(\bm a_{\lambda_1}\cdot {\bm \rho})
(\bm a_{\lambda_2}\cdot {\bm \rho})(\bm a_{\lambda_3}\cdot {\bm \rho})
-\frac{\rho^2}{4\nu_2\nu_3}[\delta_{(1)}(\bm a_{\lambda_1}\cdot{\bm \rho})+\delta_{(2)}(\bm a_{\lambda_2}\cdot{\bm \rho})
+\delta_{(3)}(\bm a_{\lambda_3}\cdot{\bm \rho})]\, \nonumber\\
&&+\frac{Q^2}{4\nu_3}[\delta_{(1)}(\bm a_{\lambda_1}\cdot{\bm \rho})+\delta_{(2)}(\bm a_{\lambda_2}\cdot{\bm \rho})
-\delta_{(3)}(\bm a_{\lambda_3}\cdot{\bm \rho})]\, \nonumber\\
&&+\frac{2i}{ S}\left[\delta_{(1)}{\bm a}_{\lambda_1}\cdot({\bm N}_1-\bm V)+
\delta_{(2)}{\bm a}_{\lambda_2}\cdot({\bm N}_2+\nu_2\bm V)+
\delta_{(3)}{\bm a}_{\lambda_3}\cdot({\bm N}_3+\nu_3\bm V)\right]\, \nonumber\\
&&+U_1
-\nu_2\,U_2-\nu_3\,U_3+2\nu_2\nu_3\,U_4\Bigg>\,\, .
\end{eqnarray}
The operator $\cal P$ in front of the integral takes the odd part of the 
integral with respect to the external vector potential ${\bm A}(\phi)$. We have also introduced the operator  
\begin{eqnarray}
[\bm X\cdot\bm Y]_{\lambda_i\lambda_j}\equiv (\bm X\cdot\bm a_{\lambda_i})
(\bm Y\cdot\bm a_{\lambda_j})+(\bm X\cdot\bm a_{\lambda_j})
(\bm Y\cdot\bm a_{\lambda_i})-(\bm X\cdot\bm Y)
(\bm a_{\lambda_i}\cdot\bm a_{\lambda_j})
\end{eqnarray}
for any two vectors $\bm X$ and $\bm Y$ and we employed the abbreviations
\begin{eqnarray}\label{U}
&&U_1=-(\bm a_{\lambda_1}\cdot {\bm \rho})[\bm d_2\cdot \bm d_3]_{\lambda_2\lambda_3}+
(\bm a_{\lambda_2}\cdot {\bm \rho})[\bm d_2\cdot \bm d_3]_{\lambda_1\lambda_3}+
(\bm a_{\lambda_3}\cdot {\bm \rho})[\bm d_2\cdot \bm d_3]_{\lambda_1\lambda_2}\nonumber\\
&&+2(\bm a_{\lambda_1}\cdot{\bm \rho})(\bm a_{\lambda_2}\cdot \bm d_2)(\bm a_{\lambda_3}\cdot\bm d_3)
-(\bm d_2\cdot{\bm \rho})[\bm d_3\cdot\bm a_{\lambda_2}]_{\lambda_1\lambda_3}
-(\bm d_3\cdot{\bm \rho})[\bm d_2\cdot\bm a_{\lambda_3}]_{\lambda_1\lambda_2}\,,
\nonumber\\
&&\nonumber\\
&& U_2=(\bm a_{\lambda_1}\cdot {\bm \rho})[\bm d_1\cdot \bm d_2]_{\lambda_2\lambda_3}-
(\bm a_{\lambda_2}\cdot {\bm \rho})[\bm d_1\cdot \bm d_2]_{\lambda_1\lambda_3}+
(\bm a_{\lambda_3}\cdot {\bm \rho})[\bm d_1\cdot \bm d_2]_{\lambda_1\lambda_2}\nonumber\\
&&+2(\bm a_{\lambda_2}\cdot{\bm \rho})(\bm a_{\lambda_1}\cdot \bm d_2)(\bm a_{\lambda_3}\cdot\bm d_1)
-(\bm d_1\cdot{\bm \rho})[\bm d_2\cdot\bm a_{\lambda_3}]_{\lambda_1\lambda_2}
-(\bm d_2\cdot{\bm \rho})[\bm d_1\cdot\bm a_{\lambda_1}]_{\lambda_2\lambda_3}\,,\nonumber\\
&&\nonumber\\
&&
U_3=(\bm a_{\lambda_1}\cdot {\bm \rho})[\bm d_1\cdot \bm d_3]_{\lambda_2\lambda_3}+
(\bm a_{\lambda_2}\cdot {\bm \rho})[\bm d_1\cdot \bm d_3]_{\lambda_1\lambda_3}-
(\bm a_{\lambda_3}\cdot {\bm \rho})[\bm d_1\cdot \bm d_3]_{\lambda_1\lambda_2}\nonumber\\
&&+2(\bm a_{\lambda_3}\cdot{\bm \rho})(\bm a_{\lambda_1}\cdot \bm d_3)(\bm a_{\lambda_2}\cdot\bm d_1)
-(\bm d_1\cdot{\bm \rho})[\bm d_3\cdot\bm a_{\lambda_2}]_{\lambda_1\lambda_3}
-(\bm d_3\cdot{\bm \rho})[\bm d_1\cdot\bm a_{\lambda_1}]_{\lambda_2\lambda_3}\,,\nonumber\\
&&\nonumber\\
&& U_4=(\bm a_{\lambda_1}\cdot \bm d_2)[\bm d_1\cdot \bm d_3]_{\lambda_2\lambda_3}+
(\bm a_{\lambda_1}\cdot \bm d_3)[\bm d_1\cdot \bm d_2]_{\lambda_2\lambda_3}-
(\bm a_{\lambda_1}\cdot \bm d_1)[\bm d_2\cdot \bm d_3]_{\lambda_2\lambda_3}\nonumber\\
&&+2(\bm a_{\lambda_1}\cdot \bm d_1)(\bm a_{\lambda_2}\cdot \bm d_2)(\bm a_{\lambda_3}\cdot \bm d_3)
+(\bm d_1\cdot \bm d_2)(\bm d_3\cdot( \delta_{(2)}\bm a_{\lambda_2}-
\delta_{(3)}\bm a_{\lambda_3})) \nonumber\\
&&-(\bm d_1\cdot \bm d_3)(\bm d_2\cdot( \delta_{(2)}\bm a_{\lambda_2}-
\delta_{(3)}\bm a_{\lambda_3}))-(\bm d_2\cdot \bm d_3)(\bm d_1\cdot( \delta_{(2)}\bm a_{\lambda_2}+
\delta_{(3)}\bm a_{\lambda_3}))\, .
\end{eqnarray}

For $\lambda_3=3$ we have

\begin{eqnarray}\label{S33final}
&&R_{\lambda_1\lambda_2\,3}^{(3)}=\frac{i\omega_1e^3}{2\pi^2}\sqrt{\frac{k_3^2}{(\varkappa k_3)^2}}\,\,
{\cal P}\int\negthickspace\negthickspace\int\limits_0^{\infty}
\negthickspace\negthickspace\int\frac{ds_1\,ds_2\,ds_3}{ S^2}\int d\phi\nonumber\\
&&\times\exp\left[i\left(\Omega\phi+\frac{4\omega_1\Omega s_2s_3}{m^2 S}-
\frac{s_1s_3}{S}Q^2+F_0- S +{\bm \rho}\cdot{\bm G}
\right)\right]\nonumber\\
&&\times\Bigg<4(\tau_1+\tau_3)({\bm a}_{\lambda_1}\cdot{\bm N}_1) ({\bm a}_{\lambda_2}\cdot{\bm N}_2)
\nonumber\\
&&+2({\bm a}_{\lambda_1}\cdot{\bm N}_1)\big[4\nu_2\tau_1({\bm a}_{\lambda_2}\cdot{\bm V})-{\bm \rho}\cdot{\bm a}_{\lambda_2}+
2\nu_2\nu_3\big({\bm a}_{\lambda_2}\cdot(\bm d_2+\bm d_3)\big)\big]
\nonumber\\
&&-2({\bm a}_{\lambda_2}\cdot{\bm N}_2)[4\tau_1({\bm a}_{\lambda_1}\cdot{\bm V})-{\bm \rho}\cdot{\bm a}_{\lambda_1}+
2\nu_3({\bm a}_{\lambda_1}\cdot\bm d_3)]
\nonumber\\
&&-4\nu_2\tau_3[\bm d_2\cdot\bm d_3]_{\lambda_1\lambda_2}-4\nu_2\tau_1[\bm d_1\cdot\bm d_2]_{\lambda_1\lambda_2}
\nonumber\\
&&+2\tau_1[\bm \rho\cdot\bm d_1]_{\lambda_1\lambda_2}+2\tau_2[\bm \rho\cdot\bm d_2]_{\lambda_1\lambda_2}+2\tau_3[\bm \rho\cdot\bm d_3]_{\lambda_1\lambda_2}
\nonumber\\
&&-\frac{\tau_2}{\nu_2}[\bm \rho\cdot\bm \rho]_{\lambda_1\lambda_2}-\frac{2i}{S}(\tau_1+\tau_3)\delta_{(3)}
\Bigg>\,\, .
\end{eqnarray}
Equations (\ref{S2final}),  (\ref{S3final}) and  (\ref{S33final}) are valid for an arbitrary vector potential $\bm A(\phi)$. The structure of Eqs. (\ref{S2final}),  (\ref{S3final}) and  (\ref{S33final}) is very similar to those obtained in the investigation of photon splitting in a laser field. The main difference is the presence of the term proportional to $Q^2$ in the exponent. However, as it will be seen below, this term is crucially important for the properties of the amplitudes.

\section{The case of a monochromatic, circularly polarized plane wave}

We consider now the laser field to be a monochromatic plane wave with frequency $\omega_0$. Then, after the integration over $\phi$ in Eqs. (\ref{S2final}),  (\ref{S3final}) and  (\ref{S33final}), we arrive at the following form for the helicity amplitude $M_{\sigma_1\sigma_2}$:
\begin{equation}
M_{\sigma_1\sigma_2}=\sideset{}{'}\sum_{n=-\infty}^\infty 2\pi\delta(\omega_2-\omega_1-n\omega_0)M_{n,\,\sigma_1\sigma_2}\,.
\end{equation}
Here the prime indicates that the summation is performed over odd numbers $n$, and $M_{n,\,\sigma_1\sigma_2}$ is the invariant amplitude of the process with absorption of $n$ laser photons if $n$ is positive or emission of $|n|$ laser photons to the laser field if $n$ is negative. Unlike the process of photon splitting into two real photons, in our case $n$ can be negative with the only restriction $n>-\omega_1/\omega_0$, following from the condition $\omega_2>0$.

The cross section corresponding to the amplitude $M_{n,\,\sigma_1\sigma_2}$ has the usual form 
\begin{eqnarray}
\label{sigma}
d\sigma_{n,\,\sigma_1\sigma_2}=|M_{n,\,\sigma_1\sigma_2}|^2\,
\frac{\omega_2\sin\theta\,d\theta\,d\varphi}{16\pi^2\omega_1}\,,
\end{eqnarray}
where $\theta$ is the angle between vectors $\bm k_2$ and $\bm k_1$,  and $\varphi$ is the azimuth angle of the vector
 $\bm k_{2\perp}$ in the plane perpendicular to $\bm \varkappa$. In terms of $\Omega=\omega_2-\omega_1$ and $\theta$ we have 
\begin{eqnarray}\label{nu23}
&&\varkappa k_2=(\omega_1+\Omega)(1+\cos\theta)\, ,  \quad
 \varkappa k_3=(\omega_1-\Omega)-(\omega_1+\Omega)\cos\theta\, ,  \nonumber\\
&& |\bm{k}_{2\perp}|=(\omega_1+\Omega)\sin\theta\, ,  \quad k_3^2=-\bm{q}^2=-2\omega_1^2(1-\cos\theta)-2\Omega^2(1+\cos\theta)\, .
\end{eqnarray}

The vector potential of a monochromatic, elliptically polarized laser field is
\begin{equation}
\bm A(\phi)=\bm A_1\cos(\omega_0\phi)+\bm A_2\sin(\omega_0\phi)\, ,
\end{equation}
with $\bm A_1\cdot\bm A_2=\bm A_1\cdot\bm\varkappa=\bm A_2\cdot\bm\varkappa=0$. In general, the vectors $\bm A_1$ and $\bm A_2$ are not parallel to the unit vectors $\bm a_1$ and $\bm a_2$ in Eq. (\ref{varepsilon}), respectively, and there is an azimuth asymmetry in the cross section for an elliptically polarized laser field (we call to mind that $\bm a_1$ has been chosen to be parallel to $\bm{k}_{2\perp}$). However, in a circularly polarized laser field which we consider below, the azimuth asymmetry is absent and the vector $\bm A_1$ can be directed along $\bm a_1$ and $\bm A_2$ along $\bm a_2$ (for positive helicity (p.\,h.)) or $\bm A_2$ along $-\bm a_2$ (for negative helicity (n.\,h.)). Here, we consider only the case of positive helicity of the laser field while the amplitudes for negative helicity can be obtained by means of the relation $M_{\sigma_1\sigma_2}(\mbox{n.\,h.})=M_{\bar\sigma_1\bar\sigma_2}(\mbox{p.\,h.})$, with $\bar\sigma_i$ denoting the helicity opposite to $\sigma_i$. Also, we set $|\bm A_1|=|\bm A_2|=A$ and $\xi=eA/m$. Using Eqs. (\ref{MviaS1}) and (\ref{MviaS2}), we represent $M_{n,\sigma_1\sigma_2}$ as
\begin{equation}\label{M23}
 M_{n,\sigma_1\sigma_2}=M^{(2)}_{n,\sigma_1\sigma_2}+M^{(3)}_{n,\sigma_1\sigma_2}\,,
\end{equation}
where for $M^{(2)}_{n,\sigma_1\sigma_2}$ we   find from Eq. (\ref{S2final})
\begin{eqnarray}\label{M2circular}
&&M^{(2)}_{n,++}=\delta_{n,-1}(G_2^{(+)}+G_3^{(+)})+\delta_{n,1}(G_1^{(-)}+G_3^{(-)})\,;\nonumber\\
&&M^{(2)}_{n,+-}=-\delta_{n,1}(G_1^{(-)}+G_2^{(-)})\,;\nonumber\\
&&M^{(2)}_{n,-+}=-\delta_{n,-1}(G_1^{(+)}+G_2^{(+)})\,;\nonumber\\
&&M^{(2)}_{n,--}=\delta_{n,-1}(G_1^{(+)}+G_3^{(+)})+\delta_{n,1}(G_2^{(-)}+G_3^{(-)})\nonumber\,;\\
&&G_j^{(\pm)}=\frac{i Z\alpha^2\rho\xi}{\nu_3\omega_1Q^2}\int_0^\infty\frac{ ds}{s}\int_0^1du
\exp\left\{-is\left[1+u(1-u)Q^2\delta_{j,3}+\xi^2\left(1-\frac{\sin^2\vartheta}{\vartheta^2}\right)\right]\right\}\nonumber\\
&&\times\mbox{e}^{i\vartheta}\left[2i\left(\mbox{e}^{i\vartheta} -   \frac{\sin\vartheta}{\vartheta}\right)+
\,\frac{\sin\vartheta}{u(1-u)}\right]\,;\quad\vartheta=\pm 2u(1-u)s\eta\nu_j \, .
\end{eqnarray}
 
In order to write $M^{(3)}_{n,\sigma_1\sigma_2}$ in a compact form, we introduce the following definitions:
\begin{eqnarray}\label{not}
&&f_j=\sin(2\eta s_j\tau_j)\mbox{e}^{i2\eta s_j\tau_j}\,,\quad 
\zeta=\frac{i}{\tau_1\eta}\left(\frac{f_2}{s_2\tau_2}-\frac{f_3}{s_3\tau_3}\right)\,,\quad \phi_0=\arg{\zeta}\,,\nonumber\\
&&F_0=\frac{\xi^2}{4\eta^2 S\tau_1\tau_2\tau_3}\left(\frac{\nu_2\nu_3}{\tau_1}|f_1|^2+\frac{\nu_2}{\tau_2}|f_2|^2
-\frac{\nu_3}{\tau_3}|f_3|^2\right)-\xi^2 S\, ,\nonumber\\
&& z=\xi|\zeta|\frac{\rho s_2s_3}{S}\,,\quad  S=s_1+s_2+s_3\, .
\end{eqnarray}
In terms of these functions, the contribution $M^{(3)}_{n,\sigma_1\sigma_2}$ has the form
\begin{eqnarray}\label{M3circular}
M^{(3)}_{n,\sigma_1\sigma_2}&=&\frac{iZ\alpha^2\rho}{\nu_3\omega_1Q^2}\int\negthickspace\negthickspace\int\limits_0^{\infty}
\negthickspace\negthickspace\int
\frac{ds_1\,ds_2\,ds_3}{ S^2}\exp\left[i\left(4n\eta\frac{s_2s_3}{S}-\frac{s_1s_3}{S}Q^2+F_0- S\right)\right]\nonumber\\
&&\times \sum_{j=-3}^{3}\, \mbox{e}^{-i(n+j)\phi_0}\,J_{n+j}(z)B_{j,\sigma_1\sigma_2}\, ,
\end{eqnarray}
where $J_l(z)$ are ordinary Bessel functions. The coefficients $B_{j,\sigma_1\sigma_2}$ are expressed via the functions $f_j$ introduced in Eq. (\ref{not}) and they are presented in the Appendix. The expressions (\ref{M2circular}) and (\ref{M3circular}) are valid for any value of the parameters $\eta$ and $\xi$. Note that the expansion of the amplitude $M_{n,\sigma_1\sigma_2}$ for a fixed $n$, contains in general all odd powers $k$ of the parameter $\xi$, starting from $k=|n|$. The terms with $k>|n|$ in $M_{n,\sigma_1\sigma_2}$ correspond to rescattering processes of absorption and emission of laser photons with a net emission or absorption of $|n|$ laser photons.

\section{Asymptotics of the amplitudes}
In this Section we consider some asymptotics of the amplitudes $M_{n,\sigma_1\sigma_2}$ in Eq. (\ref{M23}). For optical laser fields we have $\omega_0\ll m$. Also, at $\omega_1\lesssim m$ the cross section is too small in comparison with that of usual Delbr\"uck scattering in a pure Coulomb field. Therefore, we consider below the most interesting case, from an experimental point of view, $\omega_1\gg m$. In this case, the main contribution to the total cross section comes from scattering angles $\theta\lesssim m/\omega_1\ll 1$. We also assume that $\theta$ is not too small, namely $\theta\gg \theta_0=\sqrt{\omega_0/\omega_1}$. In these approximations we have
\begin{eqnarray}\label{nu23a}
&& \quad  \nu_3=\frac{\theta^2}{4}\, ,\quad
 \rho=\frac{\omega_1\theta}{m}\, ,  \quad  Q^2=\rho^2\left(1+\nu_3\right)\, .
\end{eqnarray}
Note that the amplitudes in the region $\theta\sim \theta_0$ can also be easily derived from the general expressions (\ref{M2circular}) and (\ref{M3circular}). However, it is difficult to investigate experimentally this region of very small $\theta$ and then we do not consider it here.

We start with the case of small $\eta=\omega_0\omega_1/m^2$ and large $\xi$ but fixed $\chi=\eta\xi$ and $\rho$. Nowadays, these conditions can be provided by using available photon sources and strong optical lasers \cite{Emax}. Note that this limit of parameters provides the fulfillment of the quasiclassical approximation. It is convenient to present the cross section (\ref{sigma})
in the form
\begin{eqnarray}\label{sigma1}
\label{d_sigma}
d\sigma_{n,\,\sigma_1\sigma_2}=\sigma_0\,|A_{n,\,\sigma_1\sigma_2}|^2\,
\frac{d\rho^2}{\rho^4}\, ,\quad \sigma_0=\frac{4}{\pi}\alpha^2(Z\alpha)^2\lambda_C^2\chi^2\, ,
\end{eqnarray}
where $\lambda_C=1/m$ is the Compton wavelength and the dimensionless amplitudes  $A_{n,\, ++}$ and $A_{n,\, +-}$ read:
\begin{eqnarray}\label{An}
A_{n,\,++}&=&(\delta_{n,1}+\delta_{n,-1})\frac{i}{\rho}\int_0^\infty ds\int_0^1 du\mbox{e}^{-i\psi_0}\,[1-2u(1-u)]\, \nonumber\\
&&+i\int_0^\infty ds\int_0^1 du\,u\int_0^1dy\mbox{e}^{-i\psi_1}\sum_{j=-2}^{2}(-1)^{n+j}J_{n+j}(z)\,{\cal B}_{j,\,++}\, ,\nonumber\\
A_{n,\,+-}&=&i\int_0^\infty ds\int_0^1 du\,u\int_0^1dy\mbox{e}^{-i\psi_1}\sum_{j=-2}^{0}(-1)^{n+j}J_{n+j}(z)\,{\cal B}_{j,\,+-}\, .\nonumber\\
\end{eqnarray}
Here, the functions $\psi_{0,1}$, $z$ and the coefficients ${\cal B}_{j,\,\sigma_1\sigma_2}$ are
\begin{eqnarray}
&&\psi_0=s+\frac{4}{3}u^2(1-u)^2\chi^2s^3\, ,\quad\psi_1=\psi_0+y(1-y)u^2\rho^2s\,,\quad z=4\chi\rho s^2u^2(1-u)y(1-y)\,,\nonumber\\
&&{\cal B}_{2,\,++}=2\chi s^2u^2(1-u)[1-2u(1-u)](1-2y)^2\,,\nonumber\\
&&{\cal B}_{1,\,++}=[1-2u(1-u)]\left[-\frac{2}{\rho}+4i\rho s u^2y(1-y)\right]-i\rho s u(2-3u+2u^2) \,,\nonumber\\
&&{\cal B}_{0,\,++}=2i\frac{1-u}{\chi s}+\frac{\rho^2}{2\chi}u[-1+4u(1-u)y(1-y)]-8\chi s^2u(1-u)^2(1-u+u^2)\,,\nonumber\\
&&{\cal B}_{-1,\,++}=[1-2u(1-u)]\left[-\frac{2}{\rho}+4i\rho s u^2y(1-y)\right]+i\rho s u(2-5u+6u^2-4u^3) \,,\nonumber\\
&&{\cal B}_{-2,\,++}=-{\cal B}_{2,\,++}\,,\nonumber\\
&&{\cal B}_{0,\,+-}=-\frac{2\rho^2}{\chi}u^2(1-u)y(1-y)\,,\nonumber\\
&&{\cal B}_{-1,\,+-}=4i\rho s u^2(1-u)^2\,,\nonumber\\
&&{\cal B}_{-2,\,+-}=8\chi s^2u^2(1-u)^3\,.
\end{eqnarray}
The amplitudes  $A_{n,\, --}$ and $A_{n,\, -+}$ are obtained from $A_{n,\, ++}$ and $A_{n,\,+-}$ by the substitution 
\begin{equation}
A_{n,--}=-A_{-n,++}\,,\quad A_{n,-+}=-A_{-n,+-}\,.
\end{equation}

For $\omega_1\gg m$ it is difficult to measure the energy $\omega_2$ of the final photon with an accuracy of the order of $\omega_0$ (energy of optical lasers photons). Besides, it is as well very difficult to measure the polarization of the final photons. Therefore, we consider below the cross section $d\sigma$ summed up over $n$ and over the polarization of the final photon and averaged over the polarization of the initial photon, i. e.
\begin{equation}
d\sigma=\frac{1}{2}\sum_{n,\sigma_1,\sigma_2}d\sigma_{n,\sigma_1,\sigma_2}\,.
\end{equation}

If $\rho\ll 1$ and $\chi \lesssim 1$ than the leading terms of the amplitudes
are linear in $\rho$ and the nonzero  $A_{n,\,\sigma_1\sigma_2}$  are
\begin{eqnarray}
\label{A_small_rho}
&&A_{n,++}=\rho(\delta_{n,1}-\delta_{n,-1}){\cal F}_1\,,\quad A_{n,+-}=\rho(\delta_{n,1}{\cal F}_2+\delta_{n,3}{\cal F}_3)\,,\nonumber\\
&&{\cal F}_1=\frac{2}{3}\int_{0}^{\infty} ds\int_0^1 du\, \mbox{e}^{-i\psi_0}s u^2 (1-u) [-3+4u-4u^2+4i\chi^2 s^3 u^2(1-u)^2(1-u+u^2)]\,,\nonumber\\
&&{\cal F}_2=\frac{4}{3}\int_{0}^{\infty} ds\int_0^1 du\, \mbox{e}^{-i\psi_0}s u^3 (1-u)^2 [-3+2i\chi^2 s^3 u^2(1-u)^2]\,,\nonumber\\
&&{\cal F}_3=-i\frac{8}{3}\chi^2\int_{0}^{\infty} ds\int_0^1 du\, \mbox{e}^{-i\psi_0}s^4 u^5 (1-u)^4\,.
\end{eqnarray}
Therefore, at $\rho\ll 1$
\begin{equation}\label{sigma_smallrho}
d\sigma=\sigma_0\,(2|{\cal{F}}_1|^2+|{\cal{F}}_2|^2+|{\cal{F}}_3|^2) \frac{d\rho^2}{\rho^2}\, ,
\end{equation}
and the integrated cross section contains a logarithmic divergence at small $\rho$. This divergence is eliminated by including the atomic form factor into Eq. (\ref{d_sigma}) and by replacing $\rho^4$ in Eq. (\ref{d_sigma}) by $Q^4\approx (\rho^2+4\Omega^2/m^2)^2$ with $\Omega=n\omega_0$. Thus, we obtain with the logarithmic accuracy
\begin{equation}\label{sigma_tot}
\sigma=\sigma_0\,(2|{\cal{F}}_1|^2+|{\cal{F}}_2|^2+|{\cal{F}}_3|^2)\log\left(\frac{1}{\rho_{min}^2}\right)\,, \quad \rho_{min}=\mbox{max}\left\{\frac{1}{m r_{scr}},\,\frac{\omega_0}{m}\right\}\,,
\end{equation}
where $r_{scr}\sim Z^{-1/3}/(m\alpha)$ is the screening radius corresponding to the atomic form factor. Note that it is also not difficult to obtain the integrated cross section beyond the logarithmic accuracy.  In Eq. (\ref{sigma_tot}) the functions ${\cal F}_i$ depend only on $\chi$ with ${\cal{F}}_1(\chi=0)=11/90$, ${\cal{F}}_2(\chi=0)=1/15$, and ${\cal{F}}_3(\chi=0)=0$. This dependence reflects the contributions of high order terms with respect to the laser field.

We consider now the asymptotics of the amplitudes in Eq. (\ref{An}) at $\rho\gg 1$ and $\chi\gtrsim 1$.
The leading contribution  to the cross section $d\sigma$ comes from $n=\pm 1$ so that
\begin{eqnarray}
\label{A_large_rho}
&&A_{n,++}=\frac{1}{\rho}(\delta_{n,1}-\delta_{n,-1}){\cal F}_4\,,\quad
A_{n,+-}=\frac{1}{\rho}\delta_{n,1}{\cal F}_5\,,\nonumber\\
&&{\cal F}_4=i\int_{0}^{\infty} ds\int_0^1 du\, \mbox{e}^{-i\psi_0} (1-2u+2u^2)\,,\nonumber\\
&&{\cal F}_5=4i\int_{0}^{\infty} ds\int_0^1 du\, \mbox{e}^{-i\psi_0} u (1-u)\,.
\end{eqnarray}
Therefore,  at $\rho\gg1$
\begin{equation}\label{sigma_largerho}
d\sigma=
\sigma_0\,(2|{\cal{F}}_4|^2+|{\cal{F}}_5|^2) \frac{d\rho^2}{\rho^6}\, .
\end{equation}
Finally, we also present the amplitudes at $\chi\ll 1$ and fixed $\rho$. This asymptotics corresponds to the Born approximation with respect to the laser field. The amplitudes linear in $\chi$ are
\begin{eqnarray}
\label{A_small_chi}
A_{1,++}&=&\rho\int_0^1 dy\int_0^1\frac{ du\, u^2}{{\cal D}}\Bigg\{2 y(1-y)u(1-2u+2u^2)\nonumber\\
&&+\frac{1}{{\cal D}}[4y(1-y)u^3+2-5u+6u^2-4u^3]\nonumber\\
&&-\frac{1}{{\cal D}^2}2u^2(1-u)y(1-y)\rho^2[1-4u(1-u)y(1-y)]\Bigg\}
\,,\nonumber\\
A_{-1,++}&=&\rho\int_0^1 dy\int_0^1\frac{ du\, u^2}{{\cal D}}\Bigg\{2 y(1-y)u(1-2u+2u^2)\nonumber\\
&&+\frac{1}{{\cal D}}[4y(1-y)u(2-4u+3u^2)-2+3u-2u^2]\nonumber\\
&&+\frac{1}{{\cal D}^2}2u^2(1-u)y(1-y)\rho^2[1-4u(1-u)y(1-y)]\Bigg\}
\,,\nonumber\\
A_{1,+-}&=&4\rho\int_0^1 dy\int_0^1\frac{ du\, u^3(1-u)^2}{{\cal D}^2}\Bigg[1-\frac{2\rho^2u^2y^2(1-y)^2}{{\cal D}}\Bigg]\,,\nonumber\\
A_{-1,+-}&=&8\rho^3\int_0^1 dy\int_0^1\frac{ du\, u^5(1-u)^2y^2(1-y)^2}{{\cal D}^3}\, ,\nonumber\\
&&{\cal D}=1+y(1-y)u^2\rho^2\,.
\end{eqnarray}
Note that the above amplitudes for $n=-1$ are, up to an overall factor, nothing but the amplitudes of photon splitting in a Coulomb field in the Born approximation. We have checked that our amplitudes are in agreement with the results obtained in \cite{LMMST03}.

\section{Discussion of the results and conclusion}
\begin{figure}[ht]
\begin{center}
\includegraphics[width=8cm]{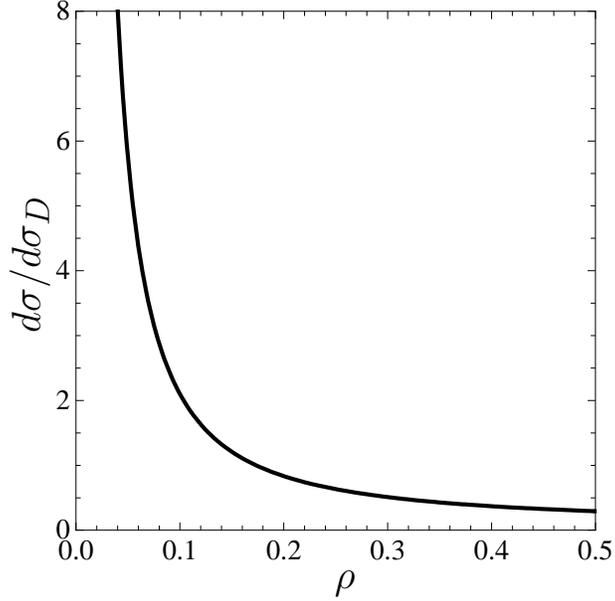}
\caption{\label{dsig/dsigD} The ratio $d\sigma/d\sigma_D$ in units $\chi^2/(Z\alpha)^2$ obtained in the Born approximation.} 
\end{center} 
\end{figure}

At $\omega_1\gg \omega_0$, the main background to the process under consideration is the usual Delbr\"uck scattering (we denote the cross section of this process as $d\sigma_D$) which  takes place without any laser field. In this case $\omega_2=\omega_1$,  and the amplitude corresponds to the exchange of an even number of Coulomb quanta. In our case, instead, $\omega_2\neq \omega_1$ and therefore there is no interference  with the usual Delbr\"uck amplitude. It is important that the dependence  of $d\sigma$ and $d\sigma_D$ on $\rho$ is very different at small as well as at large $\rho$, see Refs. \cite{LMS, LMMST03}. It is natural to compare the cross sections $d\sigma$ and $d\sigma_D$ obtained both in the Born approximation. In our case, this corresponds to the amplitudes linear in $\chi$ and in $Z\alpha$, while for Delbr\"uck scattering the amplitudes in the Born approximation are proportional to $(Z\alpha)^2$. In Fig. \ref{dsig/dsigD} we plot the ratio $d\sigma/d\sigma_D$ in units of $\chi^2/(Z\alpha)^2$. It can be seen that this ratio can be bigger than unity for the realistic parameters of the laser field. For example, for an optical ($\omega_0=1\;\text{eV}$) laser with intensity of order of $10^{23}\;\text{W/cm$^2$}$ \cite{Emax}, for an initial photon energy $\omega_1=500\;\text{MeV}$ and for $Z=47$ (silver), we obtain $\chi^2/(Z\alpha)^2\approx 1.8$.

From Fig. \ref{dsig/dsigD} one sees that the region where the $d\sigma$ mainly dominates over $d\sigma_D$ is that of small $\rho$.

\begin{figure}[ht]
\begin{center}
\includegraphics[width=8cm]{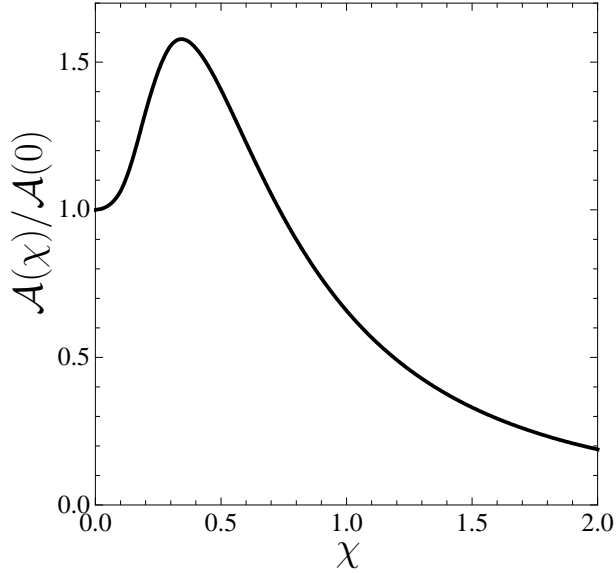}
\caption{\label{cala}  The ratio
 ${\cal A}(\chi)/{\cal A}(0)$ , where ${\cal A}(0)=139/4050$.} 
\end{center} 
\end{figure}

It is interesting to study the effect of higher order terms in $\chi$ on the small-$\rho$ differential cross section. 
This effect is determined by the $\chi$-dependence of the function ${\cal A}(\chi)=2|{\cal{F}}_1|^2+|{\cal{F}}_2|^2+|{\cal{F}}_3|^2$ (see Eq. (\ref{sigma_smallrho})).
This dependence is shown in Fig. \ref{cala}.
The function ${\cal A}(\chi)$ has the following behaviour at small and large $\chi$:
\begin{eqnarray}
{\cal A}(\chi)&=&\frac{139}{4050}+\frac{7904}{42525}\chi^2\quad \mbox{at}\; \chi\ll 1\,,\nonumber\\
{\cal A}(\chi)&=&\frac{3^{1/3}\,5\,\Gamma^6(2/3)}{4^{1/3}\, 378\, \Gamma^2(4/3)}\chi^{-8/3} \quad \mbox{at}\; \chi\gg 1\,.
\end{eqnarray}
Therefore, the high order terms in $\chi$ suppress, in comparison with the Born result, the cross section at large $\chi$ and increase it at small $\chi$, making a substantial modification already at $\chi\lesssim 1$.

\begin{figure}[ht]
\begin{center}
\includegraphics[width=10cm]{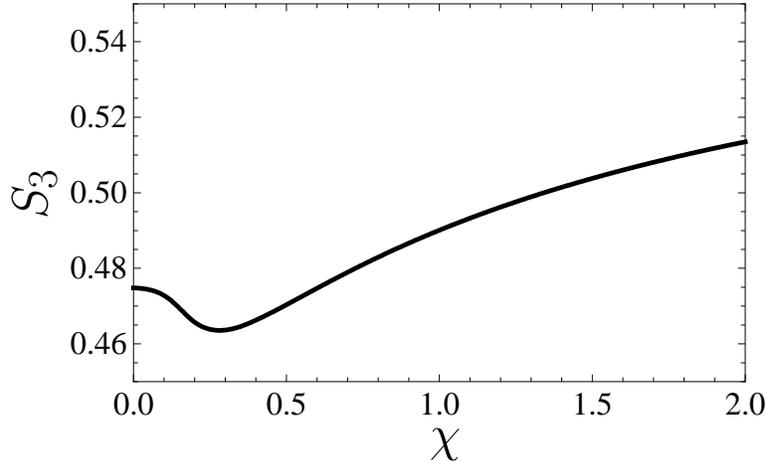}
\caption{\label{xi3} The Stokes parameter $S_3$ at small $\rho$ as a function of $\chi$.} 
\end{center} 
\end{figure}

If the initial photon is not polarized, partial polarization of the final photon in the scattering plane is produced. The corresponding Stokes parameters are \cite{Landau_b_4_1982}
\begin{eqnarray}
{S_3}&=&\frac{2\sum_n\mbox{Re}(M_{n,++}M_{n,+-}^*)}{\sum_n (|M_{n,++}|^2+|M_{n,+-}|^2)}\,,\quad S_1=S_2=0\, .
\end{eqnarray}
For $\rho\ll 1$ we find  that $S_3$ is $\rho$-independent and it is
\begin{eqnarray}
{S_3}&=&\frac{2\mbox{Re}({\cal F}_1{\cal F}_2^*)}{ 2|{\cal{F}}_1|^2+|{\cal{F}}_2|^2+|{\cal{F}}_3|^2}\,,
\end{eqnarray}
unlike the case of pure Delbr\"uck scattering where $S_3$ decreases at small $\rho$. The dependence of $S_3$ on $\chi$ at small $\rho$ is shown in Fig. \ref{xi3}. It can be seen that the polarization degree is not small and the $\chi$-dependence is smooth.

It is also interesting to compare the cross section obtained here with that of photon splitting in a pure Coulomb field, see Refs. \cite{LMS,LMMST03}. The ratio of these two cross sections can be estimated as $\chi^2/\alpha$. Therefore, for realistic parameters of laser fields and photon sources this ratio can be substantially larger than unity.

In conclusion, we have derived the Delbr\"uck scattering  amplitudes in a Coulomb field in the presence of a laser field.  The amplitudes have been calculated in the Born approximation with respect to the Coulomb field and exactly in the parameters of the laser field having arbitrary strength, spectral content and polarization. The most interesting case of a high energy initial photon  is investigated in detail for a monochromatic circularly polarized laser field. It is shown that the angular distribution of the final photon in the process under discussion substantially differs from that for Delbr\"uck scattering in a pure Coulomb field. Moreover, the value of the cross section of the process at hand may exceed that of Delbr\"uck scattering in a pure Coulomb field at realistic laser parameters and this essentially simplifies the possibility of the experimental observation of the phenomenon. Moreover, the effect of high order terms in $\chi$ is quantitatively important already at relatively small $\chi$, making the experimental investigation of these effects very attractive.

\section*{Acknowledgments}
A. I. M. gratefully acknowledges the hospitality and the financial support he has received during his visit at Max-Planck-Institute 
for Nuclear Physics. The work was supported in part by the RFBR under grant 06-02-04018.

\appendix*
\section{Coefficients for the helicity amplitudes}
Here we present the coefficients $B_{j,\sigma_1\sigma_2}$ used in Eq. (\ref{M3circular}). We introduce the abbreviations
\begin{eqnarray}
&&x_j=\frac{s_j}{ S}\, ,\quad d_1=\frac{(\tau_1-\tau_2)f_2}{2\tau_1\tau_2}+\frac{(\tau_1+\tau_3)f_3}{2\tau_1\tau_3}\,,
\quad  d_2=\frac{(\tau_2-\tau_1)f_2}{2\tau_1\tau_2}+\frac{(\tau_1-\tau_3)f_3}{2\tau_1\tau_3}\,, \nonumber\\ 
&& d_3=\frac{(\tau_2+\tau_1)f_2}{2\tau_1\tau_2}-\frac{(\tau_1+\tau_3)f_3}{2\tau_1\tau_3}\,,  
\quad h=d_1+d_2+d_3=\frac{(\tau_2+\tau_1)f_2}{2\tau_1\tau_2}+\frac{(\tau_1-\tau_3)f_3}{2\tau_1\tau_3}\, ,\nonumber\\
&&g_1=\frac{h}{\eta S}-1-ih\,,\quad g_2=\frac{h}{\eta S}-1-2if_2+i\nu_2h\,,\quad g_3=\frac{h}{\eta S}-1-2if_3+i\nu_3h\,.
\end{eqnarray}
Then, the nonzero coefficients for the contribution  $M^{(3)}_{n,++}$  are 
\begin{eqnarray}\label{++}
B_{2,++}&=&-4\rho\,\xi^2x_1\left(g_2 g_3+\nu_2\nu_3d_1^2 \right)\, ,\nonumber\\
B_{1,++}&=&-4\xi^3\left[g_1^* g_2 g_3+\nu_2\nu_3 g_1^*d_1^2+(\nu_3g_2+\nu_2g_3)|h|^2-i\nu_2\nu_3h^*(h^2+d_1^2)\right]\nonumber\\
&&+\rho^2\xi\left[\frac{(8x_1x_2-1)g_2}{\nu_3}+\frac{4x_1x_3 g_3}{\nu_2}+i\frac{\nu_2}{\nu_3}h+4ix_1 d_1\right]\nonumber\\
&&+\frac{4i\xi}{S}(g_2+g_3)+\frac{\xi}{\nu_3}Q^2\{-4(\tau_1+\tau_3)(g_2-i\nu_2 h)x_1\nonumber\\
&&-4i\nu_2x_1[2\tau_1h+\nu_3(2d_1+d_2+d_3)]+g_2-i\nu_2h\}\, ,\nonumber\\
&&\nonumber\\
B_{0,++}&=&4\rho\xi^2\Bigg[\frac{2x_2}{\nu_3}g_1^*g_2+\frac{x_3}{\nu_2}g_1^*g_3-x_1g_2g_3^*
+i(g_1^*d_1-g_2d_3^*)  +h^*d_1-\nu_2hd_3^*\nonumber\\
&&+\left(x_1\nu_2\nu_3+\frac{2\nu_2x_2}{\nu_3}+\frac{\nu_3x_3}{\nu_2}\right)|h|^2\Bigg]\nonumber\\
&&-\frac{\rho^3}{\nu_2\nu_3}[x_1(8x_2x_3-1)-x_3]+\frac{4i\rho}{ S}\left(x_1-\frac{2x_2}{\nu_3}-\frac{x_3}{\nu_2}\right)\nonumber\\
&&+\frac{1}{\nu_3}\frac{Q^2}{\rho}\left\{4(\tau_1+\tau_3)\left[-\xi^2(g_1^*-ih^*)(g_2-i\nu_2h)
+\rho^2\frac{x_1x_3}{\nu_2}\right]\right.\nonumber\\
&&-4i\nu_2\xi^2(g_1^*-ih^*)[2\tau_1h+\nu_3(2d_1+d_2+d_3)]\nonumber\\
&&\left.-4i\xi^2(g_2-i\nu_2h)[2\tau_1h^*+\nu_3(d_1^*+d_2^*)]-\rho^2\left(x_1\nu_3+\frac{\tau_2}{\nu_2}\right)+\frac{4i}{S}(\tau_1+\tau_3)\right\}\,,\nonumber\\
B_{-1,++}&=&4\xi^3\left[-g_1^*g_2g_3^*+\nu_2(\nu_3g_1^*-g_3^*)|h|^2+\nu_3g_2d_3^{*2}-i\nu_2\nu_3h(h^{*2}+d_3^{*2})\right] \nonumber\\
&&+\frac{\rho^2\xi}{\nu_2}\left(4x_1x_3 g_3^*-\frac{8x_2x_3-1}{\nu_3}g_1^*+4ix_3d_3^*-i\frac{h^*}{\nu_3}\right)+\frac{4i\xi}{S}(g_1^*+g_3^*)\nonumber\\
&&+\frac{\xi}{\nu_3}Q^2\left\{\left[\frac{4x_3(\tau_1+\tau_3)}{\nu_2}-1-\nu_3\right](g_1^*-ih^*)+\frac{4ix_3}{\nu_2}[2\tau_1h^*+\nu_3(d_1^*+d_2^*)]\right\}\,,\nonumber\\
B_{-2,++}&=&\frac{4\rho\xi^2x_3}{\nu_2}\left(g_1^*g_3^*-\nu_3d_3^{*2}\right)\, .
\end{eqnarray}
The nonzero coefficients for the contribution  $M^{(3)}_{n,+-}$  are
\begin{eqnarray}\label{+-}
B_{1,+-}&=&-\frac{\rho^2\xi}{\nu_2}[(4x_1x_3-1)g_3+i\nu_3 h]\, ,\nonumber\\
B_{0,+-}&=&\rho^3x_2\frac{8x_1x_3-1}{\nu_2\nu_3}\nonumber\\
&&-4\rho\xi^2\left[g_3\left(\frac{x_3 g_1^*}{\nu_2}-x_1g_2^*\right)
-i(g_3-i\nu_3h)d_2^*+\nu_3\left(\frac{x_3}{\nu_2}+x_1\nu_2\right)|h|^2\right]\nonumber\\
&&-\frac{4i\rho}{S}\left(x_1-\frac{x_3}{\nu_2}\right)-\frac{\rho}{\nu_3}Q^2\left[4(\tau_1+\tau_3)\frac{x_1x_3}{\nu_2}+\nu_2x_1+\frac{x_3}{\nu_2}-1\right]\,,\nonumber\\
&&\nonumber\\
B_{-1,+-}&=&\rho^2\xi\Bigg[\frac{8x_2x_3-1}{\nu_2\nu_3}g_1^*-\frac{8x_1x_2-1}{\nu_3}g_2^* -\frac{4x_1x_3-1}{\nu_2}g_3^* \nonumber\\
&&-2i(1-2x_1)d_1^*+\frac{2i(1-2x_3)}{\nu_2}d_3^*+\frac{2i(1-4x_2)}{\nu_3}d_2^*\Bigg]\nonumber\\
&&-4\xi^3\left[-g_1^*g_3g_2^*+\nu_3(\nu_2g_1^*-g_2^*)|h|^2+\nu_2g_3d_2^{*2}-i\nu_2\nu_3h(h^{*2}+d_2^{*2})\right] \nonumber\\
&&-\frac{4i\xi}{S}(g_1^*+g_2^*)+\frac{\xi}{\nu_3}Q^2\left\{-4(\tau_1+\tau_3)\left[\frac{x_3(g_1^*-ih^*)}{\nu_2}-x_1(g_2^*+i\nu_2h^*)\right]\right.\nonumber\\
&&+(1+\nu_3)(g_1^*-ih^*)-4ix_1\nu_2[2\tau_1h^*+\nu_3(2d_1^*+d_2^*+d_3^*)]\nonumber\\
&&-g_2^*-i\nu_2h^*-\frac{4ix_3}{\nu_2}[2\tau_1h^*+\nu_3(d_1^*+d_2^*)]\nonumber\\
&&+2i[2\tau_1(d_2^*+d_3^*)+(2\tau_2-\nu_2)(d_1^*+d_3^*)+2\tau_3(d_1^*+d_2^*)]\Bigg\}\,,\nonumber\\
B_{-2,+-}&=&4\rho\xi^2\Bigg[ x_1g_2^*g_3^*- g_1^*\left(\frac{2x_2}{\nu_3}g_2^*+\frac{x_3}{\nu_2}g_3^*\right)+ig_1^*d_1^*+ig_2^*d_3^*
+ig_3^*d_2^*\nonumber\\  
&&+x_1\nu_2\nu_3d_1^{*2}+\frac{2x_2\nu_2}{\nu_3}d_2^{*2}+\frac{x_3\nu_3}{\nu_2}d_3^{*2}+\nu_2d_1^*d_2^*
+\nu_3d_1^*d_3^* -d_2^*d_3^* \Bigg]\nonumber\\
&&+\frac{4\xi^2}{\rho\nu_3}Q^2\{(g_1^*-ih^*)[(\tau_1+\tau_3)(g_2^*+i\nu_2h^*)-2i\nu_2\tau_1h^*-i\nu_2\nu_3(2d_1^*+d_2^*+d_3^*)]\nonumber\\
&&+i(g_2^*+i\nu_2h^*)[2\tau_1h^*+\nu_3(d_1^*+d_2^*)]+2\nu_2(d_1^*+d_3^*)[\tau_1(d_2^*+d_3^*)+\tau_3(d_1^*+d_2^*)]\}\,,\nonumber\\
&&\nonumber\\
B_{-3,+-}&=&4\xi^3( g_1^*g_2^*g_3^*+\nu_2\nu_3g_1^*d_1^{*2}-\nu_3g_2^*d_3^{*2}-\nu_2g_3^*d_2^{*2}
+2i\nu_2\nu_3d_1^*d_2^*d_3^* )\, .
\end{eqnarray}
Finally, the coefficients for the amplitudes  $M^{(3)}_{n,-+}$ and $M^{(3)}_{n,--}$ can be obtained with the help of the substitutions
\begin{eqnarray}
&&B_{j,-+}=B_{-j,+-}^*(S\to - S)\, ,\quad B_{j,--}=B_{-j,++}^*( S\to - S)\,,
\end{eqnarray}
where the replacement $S\to- S$ means that after complex conjugation, it is also necessary to change the sign of the terms containing $S$ in Eqs. (\ref{++}) and (\ref{+-}).

\end{document}